# Revealing the EuCd$_2$As$_2$ Semiconducting Band Gap via n-type La-Doping

*Ryan A. Nelson[1], Jesaiah King[1], Shuyu Cheng[2], Archibald J. Williams[1], Christopher Jozwiak[8], Aaron Bostwick[8], Eli Rotenberg[8], Souvik Sasmal[5], I-Hsuan Kao[5], Aalok Tiwari[5], Natalie R. Jones[1], Chuting Cai[6], Emma Martin[7], Andrei Dolocan[9], Li Shi[6,7], Roland Kawakami[2], Joseph P. Heremans[2-4], Jyoti Katoch[5], Joshua E. Goldberger[1,4]\**

[1]Department of Chemistry and Biochemistry, The Ohio State University, Columbus, Ohio 43210, USA

[2]Department of Physics, The Ohio State University, Columbus, Ohio 43210, USA

[3]Department of Mechanical and Aerospace Engineering, The Ohio State University, Columbus, Ohio 43210, USA

[4]Department of Materials Science and Engineering, The Ohio State University, Columbus, Ohio 43210, USA

[5]Department of Physics, Carnegie Mellon University, Pittsburgh, PA, 15213, USA

[6]Materials Science and Engineering Program, The University of Texas at Austin, Austin, Texas 78712, USA

[7]Department of Mechanical Engineering, The University of Texas at Austin, Austin, Texas 78712, USA

[8]Advanced Light Source, E.O. Lawrence Berkeley National Laboratory, Berkeley, California 94720, USA

[9]Texas Materials Institute, The University of Texas at Austin, Austin, Texas 78712, USA

KEYWORDS

Weyl Semimetal, Topological Materials, Pnictides, Magnetic Semiconductor, ARPES

ABSTRACT

$EuCd_2As_2$ has attracted considerable interest as one of the few magnetic Weyl semimetal candidate materials, although recently there have been emerging reports that claim it to have a semiconducting electronic structure. To resolve this debate, we established the growth of n-type $EuCd_2As_2$ crystals, to directly visualize the nature of the conduction band using angle resolve photoemission spectroscopy (ARPES). We show that La-doping leads to n-type transport signatures in both the thermopower and Hall effect measurements, in crystals with doping levels at 2-6 x $10^{17}$ $e^-$ $cm^{-3}$. Both p-type and n-type doped samples exhibit antiferromagnetic ordering at 9 K. ARPES experiments at 6 K clearly show the presence of the conduction band minimum at 0.8 eV above the valence band maximum, which is further corroborated by the observation of a 0.71 – 0.72 eV band gap in room temperature diffuse reflectance absorbance measurements. Together these findings unambiguously show that $EuCd_2As_2$ is indeed a semiconductor with a substantial band gap and not a topological semimetal.

INTRODUCTION

The realization of materials with topologically non-trivial electronic structures has been a major thrust in solid-state research over the past decade. One such desired phase is the magnetic Weyl semimetal, wherein the band structure near the Fermi level is characterized by a pair of nodal band crossings with linear dispersions[1] on top of underlying magnetic order, enabling phenomena such as the chiral anomaly and its thermal analogue.[2-4] The chiral anomaly and thermal chiral anomaly feature large, nonsaturating negative electrical magnetoresistances, and thermal magnetoresistances, and thus hold great promise in electronic devices and thermal switching applications. However, efficient utilization of these transport phenomena requires materials in which the Fermi level is coincident with the nodal band crossings, and that there are no other topologically trivial bands at this energy. Almost all magnetic Weyl semimetal candidates have either trivial or non-trivial bands near the topologically non-trivial nodal crossing, as is the case for such materials as $Co_2MnGa$, $Mn_3Sn$, and $EuGa_4$.[5-7] Alternatively, $MnBi_2Te_4$ is a well known magnetic Weyl semimetal candidate that has the ideal electronic structure, but is hindered by the challenge in growing single crystals with experimentally relevant thickness in the cross-plane direction due to it being a peritectic phase with a <10 °C window of synthetic accessibility.[8,9] The identification and development of magnetic Weyl semimetals remains an important frontier in materials research.

Along these lines, $EuCd_2As_2$ has been proposed as an attractive magnetic Weyl semimetal candidate due to its ideal band structure and robust synthetic accessibility.[10] From 2016-2023 most theoretical and experimental studies of $EuCd_2As_2$ assumed that this material was semimetallic at room temperature and below, until forming a topological antiferromagnet below the Néel temperature ($T_N$) of 9.5 K with nodal crossings very close to the Γ point.[10-19] The Weyl semimetal state was thought to be accessible by forcing the structure to be ferromagnetic with polarization along the c-axis via a modest magnetic field (~1.5 T).[10, 20] The semimetallic nature was supported using a broad range of measurements. First, most in-plane resistivity measurements on single crystals remained below 100 mΩ cm (or < $10^{-3}$ Ω m), often varying between about 5 to 50 mΩ cm and increasing slightly with temperature.[10, 14, 18, 21-24] Both thermopower and Hall measurements indicated p-type doping[12, 14, 16, 19, 24-29] with carrier concentrations that depended on the synthesis method, and ranged from about $10^{15}$ cm$^{-3}$ to $10^{18}$ cm$^{-3}$ in Sn flux to about $10^{20}$ cm$^{-3}$ in salt (KCl/NaCl) flux.[24, 29-31] The variation in hole concentrations is attributed to the differing levels of Eu vacancies.[13, 24] Angle resolved photoemission spectroscopy (ARPES) corroborated the inherent p-type doping, as the Fermi level of all materials was within the valence band.[12, 13, 18, 19, 24, 30, 32, 33] This hindered direct observation of the proposed semimetallic nature, namely the presence of the conduction band minimum (CBM) that was expected to dip below the valence band maximum (VBM) at the Γ point by ~50 meV from calculations.[12, 18, 19] One previous ARPES study reported the presence of the CBM within 20 meV of the Fermi level, although the fact that this is close to the reported 5-30 meV energy resolution of the ARPES experiment introduces some ambiguity to this assignment.[19] The clear identification of the VBM and CBM by ARPES measurements of n-type samples would provide more compelling evidence of the semimetallic nature. Finally, optical reflectivity experiments show the presence of a low energy plasma frequency below 350 cm$^{-1}$ and approach unity at the lowest energy, further supporting metallic behavior in these samples.[21, 34] Low-temperature magnetoinfrared spectroscopy measurements also showed an increase in reflectivity upon application of a magnetic field below $T_N$, along with the presence of field-dependent van Hove singularities, which together were attributed to the excitations between the overlapping CBM and VBM levels whose energies have a strong magnetic field dependence in a Weyl semimetal.[35]

Recently, Santos-Cottin *et al.* reported that $EuCd_2As_2$ is actually a 0.77 eV semiconductor based on the growth of $EuCd_2As_2$ using a higher purity Eu metal source (4N) and a two-step seeded synthesis.[33] These higher purity $EuCd_2As_2$ crystals showed an insulating temperature dependence of resistivity with a 300 K value of around 3 x $10^3$ mΩ cm that rose over 6 orders of magnitude upon cooling to 10 K. Similarly high purity crystals have since been grown by other groups, and comparable insulating behavior has been observed.[16, 17, 28, 29] It is important to point out that resistivity measurements alone are subject to possible error through the presence of nonohmic or insulating contact barriers. The insulating behavior was further indicated by optical reflectivity measurements, in which a low energy reflectivity value of ~0.4 was observed in these lower doped samples, along with a strong absorption edge at 0.77 eV. Finally, pump-probe ARPES measurements that collected photoemission of the excited state from the 001 surface (using 37 eV probe excitation) indicated the presence of a long-lived state at 0.76 eV above the VBM, that was attributed to excitation to the CBM.[33] There was also a short-lived excitation at 0.13 eV above the Fermi level which was attributed to excitation to the VBM and subsequent thermalization back to the Fermi level, implying appreciable p-type doping. It's important to point out that in density functional theory calculations of the Weyl semimetal electronic structure of $EuCd_2As_2$, the VBM and CBM cross within 0.1 Å$^{-1}$ of Γ to give the Weyl nodes and the predicted ~50 meV overlap whereas the difference in energy between the VBM and CBM at the A point (0 0 π/c) is ~0.7 eV (Fig. S1).[12, 18, 19] Considering the large number of experiments that previously suggested $EuCd_2As_2$ to be semimetallic, these results along with all previous studies require a careful scrutiny of measurements and samples.

There is clearly extensive debate about the nature of the electronic structure of $EuCd_2As_2$. Accurate characterization of a narrow band gap semiconductor is often complicated by the presence of dopants or impurities. Consider the history of InN, for which early reports assigned the band gap as 1.9 eV.[36] This value was widely accepted until photoluminescence experiments revealed a true value of 0.7 eV.[37, 38] Furthermore, density functional theory predictions of the electronic structure of materials comprised of unpaired *d*- or *f*-electrons have been notoriously difficult to model accurately, due to the strong electron-electron correlations.[39] Because of the ambiguity inherent in all of the previously described measurements, the most definitive signature that would identify $EuCd_2As_2$ as a semiconductor would be static ARPES measurements on an n-doped crystal at precisely the Γ point as this is where the CBM and VBM occur, which is what we report here.

Here, we establish that $EuCd_2As_2$ is indeed a semiconductor with a 0.7 - 0.8 eV band gap, through the preparation and extensive characterization of n-type $EuCd_2As_2$ crystals. We show that the incorporation of La into the Sn-flux synthesis leads to n-type thermopowers, and Hall effect signatures. ARPES measurements at 6 K of potassium-dosed samples leads to the unambiguous detection of the CBM at 0.8 eV above the VBM, with no other states observed within the gap. Room temperature diffuse reflectance absorbance measurements further confirm an optical band gap in both p-type and n-type samples around 0.71 eV. Overall, these experiments provide strong evidence that $EuCd_2As_2$ is indeed a semiconductor.

METHODS

Single crystals with lengths >3 mm, and thicknesses > 0.5 mm were synthesized using the previously reported Sn-flux synthesis procedure.[13, 40] Elements with purity greater-than or equal-to 99.99% were mixed in a Eu:Cd:As:Sn ratio of 1:2:2:10 inside Canfield crucibles and slowly heated to 900 °C over 24 hr.[41] To grow n-type crystals, 1-10 mol% La was added to the flux. It is important to note that the actual concentration of La that incorporates into the single crystals is much less than 1% and the unincorporated La remains in the Sn flux. The reaction mixture was held at 900 °C for 20 hours to allow full reagent diffusion before slowly cooling to 500 °C at 2 °C per hour to crystalize the $EuCd_2As_2$ product. The molten flux was decanted via centrifugation at 500 °C to collect the product crystals. Sn flux was further removed by fine sanding of the surfaces.

Thermoelectric transport measurements of $EuCd_2As_2$ single crystals were conducted along the in-plane direction. Measurements from 80 K to 400 K were conducted using a Janis SuperTran cryostat system (Lakeshore) with liquid nitrogen as the cryogen. The maximum field used with this instrument was ±1.4 T. Single crystals were mounted in cryostat measurement devices constructed according to the described device schematic using conductive silver-based epoxy (Epotek H20E). Brass sheets were used as current spreaders, providing uniform electrical and thermal current flow. An insulating alumina baseplate acted as the mounting plate and heatsink. Conductive epoxy was also used to secure measurement leads directly to the crystals. The measurement leads were 20-micron copper wires. The type-T thermocouples were made using the same 20-micron copper wire as well as Constantan wire. For electrical transport measurements below 80 K a cryogen-free Quantum Design Physical Properties Measurement System (PPMS) was employed. The maximum field for this instrument was ±14 T. PPMS measurement devices were constructed in much the same way as the cryostat devices but did not include a resistive heater or thermocouples. These devices were then mounted and wired to a resistivity puck (LakeShore). No thermal measurements were performed on the QD PPMS given the absence of a thermal transport accessory.

The synchrotron-based ARPES measurements were performed at beamline 7.0.2 "MAESTRO" of the Advanced Light Source (ALS) at Berkeley National Lab. The photoelectrons are collected using a Scienta Omicron R4000 hemispherical electron analyzer, which provides energy and momentum resolution better than 30 meV and 0.01 Å$^{-1}$, respectively. The EuCd$_2$As$_2$ bulk crystal samples were cleaved in-situ and measured at a base temperature of 6 K. A full reciprocal space map using a photon energy scan was performed before potassium dosing to determine the high-symmetry planes in the momentum space, and 130 eV photon energy was selected to access the plane. The effective electron concentration induced by potassium dosing is estimated following the methods described by Katoch et al. and the details are presented in the SI.[42]

Magnetic susceptibility and magnetization data was collected via an MPMS3 superconducting quantum interference device (SQUID) magnetometer by Quantum Design. Magnetic susceptibility data acquisition occurred over a temperature range from 2 K to 300 K. Small crystals (<5 mg) of EuCd$_2$As$_2$ were encapsulated by Kapton tape and affixed to the desired field orientation of measurement in a plastic straw by another piece of Kapton tape. After cooling to 2 K with no applied field, the zero field cooled magnetic susceptibility measurements were conducted under an applied field of 500 Oe all the way to 300 K. Then, the sample was returned to 2 K under the same 500 Oe measurement field and magnetic susceptibility measurements were again conducted under a 500 Oe field up to 300 K. The magnetization vs. field data was collected at 2 K by cycling the applied field in 1000 Oe increments, starting from 0 T, between +4 T and -4 T two times. The crystals used for magnetic characterization where cleaved as small fragments of larger single crystals so as not to over-range the SQUID magnetometer. Thus, the in-plane and cross-plane crystal dimensions were comparable and no self-demagnetization correction factor was applied to the data.

Infrared spectroscopy was collected using a PerkinElmer Frontier dual-range IR spectrometer under a nitrogen atmosphere. The infrared reflectance spectra were acquired in diffuse reflectance infrared Fourier transform spectroscopy mode (DRIFTS). All material was diluted in KBr that was flame-dried in a quartz ampoule and the same salt was used as the background. Tauc-Davis-Mott relations were applied to the reflectance data assuming an allowed direct transition and three-dimensional density of states.

Time-of-flight secondary ion mass spectrometry (ToF-SIMS) was performed in the Texas Materials Institute at the University of Texas in Austin. The data collected under negative polarity used a 1 kV $O_2^+$ ion beam with a beam current of about 41 nA. The raster scanned areas were 300 μm x 300 μm. The depth profile data was collected sequentially using a 30 keV $Bi^+$ analysis beam with a raster scanned area equal to 100 μm x 100 μm centered within the sputtered area. The analysis beam current was 0.3 pA. Positive polarity data was collected in the same way except that the sputtering beam was a 0.5 kV $Cs^+$ ion beam with a similar 40 nA beam current. Both p-type and n-type $EuCd_2As_2$ samples were sputtered and measured using ToF-SIMS. The sputtering depth varied from about 1 – 6 μm.

RESULTS AND DISCUSSION

The debate about the nature of $EuCd_2As_2$ is centered on whether there is band overlap between the CBM and VBM allowing for a Weyl semimetal state to exist under ferromagnetic alignment of the spins (Figure 1a), or if it is indeed a trivial semiconductor (Figure 1b). Herein we seek to answer that question conclusively by growth and ARPES characterization of $EuCd_2As_2$ single crystals. Specifically, ARPES measurements on samples in which the conduction band is clearly visible (i.e. n-doped) without the ambiguity inherent in a pump-probe experiment offer the most compelling answer to this question. Photo-ARPES requires photoexcitation to populate and visualize the states in the conduction band, whereas visualizing the conduction band in a ground state via n-type doping after confirming the location along kz is a more precise observation. To accomplish this EuCd$_2$As$_2$ crystals were synthesized in Sn flux using the highest purity reagents commercially available (4N or greater) which resulted in the growth of batches of crystals that were either p-type or n-type with carrier densities on the order of $10^{17}$ cm$^{-3}$ at room temperature based on Hall measurements. Attempts to identify the ppm-level dopant using time-of-flight secondary ion mass spectrometry (ToF-SIMS) revealed no significant compositional differences in the p-type and n-type samples (Fig. S2,3). This likely indicates that the dopant concentration is below the ToF-SIMS detection level which is typically on the order of ppm. Indeed, the n-type carrier density suggests a dopant concentration on the order of 30 ppm. Regardless, n-type doping was reproduced in multiple growths, upon doping with La. In contrast crystals grown with >1% Li, Na, Se, Gd, Dy, Sm, or Yb added to the flux led to p-type transport signatures.

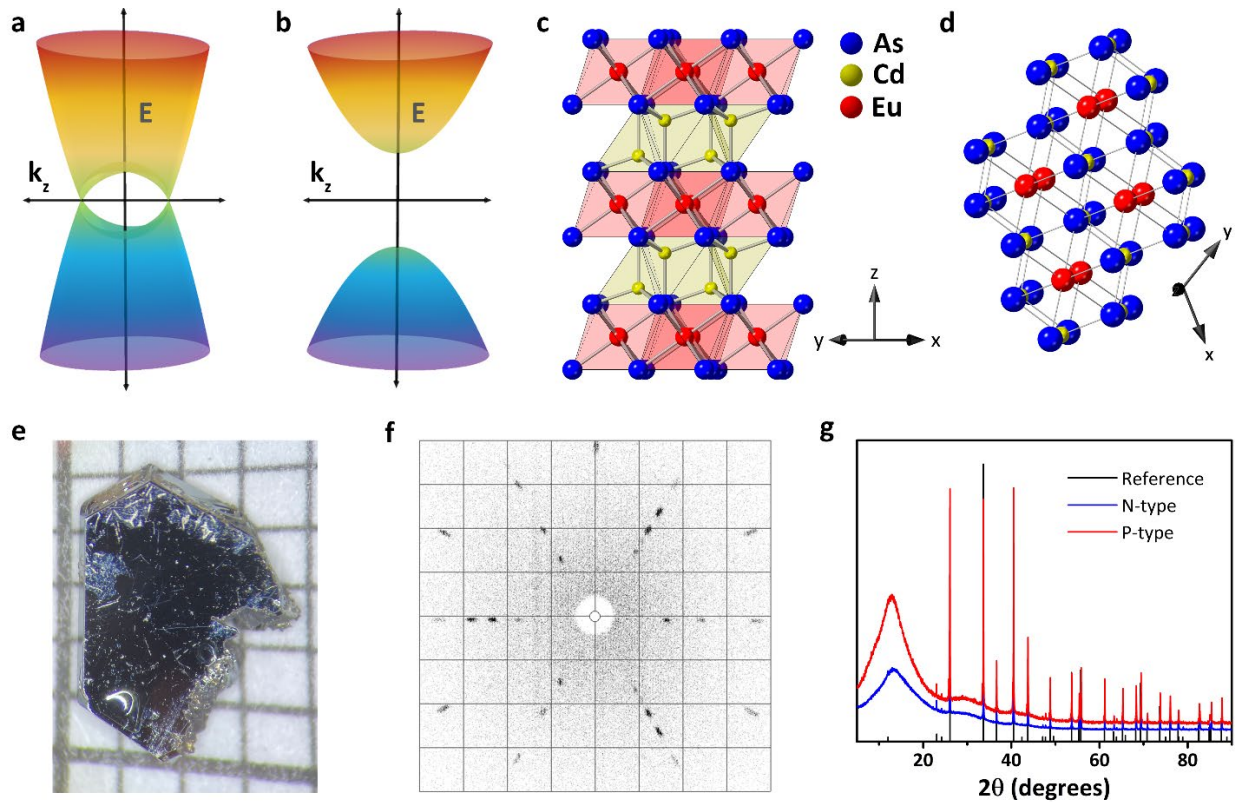

Figure 1. (a) Qualitative three-dimensional band structure of an ideal Weyl semimetal with linear dispersion near the Weyl nodes lying along the $k_z$ axis. (b) Qualitative three-dimensional band structure of a trivial direct-gap semiconductor. In schematics (a) and (b) the $k_z$ directions lie horizontally. In (a) the energy gap in the zone center has been reported to be ~50 meV, whereas in (b), the energy gap in the zone center is reported to be 0.8 eV. (c) Crystal structure of $EuCd_2As_2$ as viewed along the 110-axis with the 001-axis aligned vertically. The yellow highlighted layers represent the edge-sharing $CdAs_4$ tetrahedra sandwiched between the layers of edge-sharing $EuAs_6$ octahedra in red. (d) Crystal structure of $EuCd_2As_2$ as viewed along the 001-axis to emphasize the six-fold rotational symmetry. (e) Image of a ~4 x 3 mm$^2$ p-type $EuCd_2As_2$ crystal with the 001-axis perpendicular to the plane of the page. (f) Laue back-scattered diffraction pattern measured along the 001-axis of a p-type $EuCd_2As_2$ sample showing the six-fold rotational symmetry. (g) Powder X-ray diffraction of ground up crystals of p-type and n-type $EuCd_2As_2$. The large hump at low angles is a background feature caused by reflection from the low-volume acrylic sample holder.

$EuCd_2As_2$ has a layered crystal structure comprised of layers of edge-sharing $CdAs_4$ tetrahedra, sandwiched between single-octahedron-thick layers of edge-sharing $EuAs_6$ octahedra (Fig. 1c). Using an ionic bonding formality, the Eu and Cd oxidation states are both $2^+$ while that of the As is $3^-$. The trigonal symmetry becomes clear by viewing the structure along the 001-axis (Fig. 1d) and is manifested in the crystal growth shape (Fig. 1e) and Laue backscattered diffraction pattern (Fig. 1f). Phase confirmation by powder x-ray diffraction shows that both p-type and n-type $EuCd_2As_2$ retain the expected crystal structure (Fig. 1g). The lattice parameters were refined to values of a = 4.43965(9) Å and c = 7.32665(8) Å for p-type $EuCd_2As_2$ and refined to a = 4.44015(11) Å and c = 7.32667(11) Å for n-type $EuCd_2As_2$ (Fig. S4). These values are in excellent agreement with previously reported values and show no significant structural difference between the p- and n-type samples.[10, 11, 14, 16, 17, 21, 28, 33, 43-47]

Thermoelectric transport properties were measured using both a Janis SuperTran cryostat system (Fig. 2a) and a Quantum Design PPMS. The temperature dependent resistivities obtained for all samples of EuCd2As2 are similar in magnitude to previous reports on the metallic samples (Fig. 2b). The p-type resistivity shows a sharp cusp at the magnetic transition temperature ($T_N \cong 9.2$ K), peaking just under 60 mΩ cm and ranging from about 29 mΩ cm at 2 K to nearly 100 mΩ cm at 400 K. The cusp at the magnetic transition at the magnetic transition temperature has been observed in nearly all previous p-type transitions and arises due to carrier scattering with the spin-waves that emerge below this temperature. The n-type resistivities, however, show an absence of a cusp at the magnetic transition. Instead, the n-type samples although close in magnitude, have varying temperature coefficients of resistivities. One crystal (crystal 1) had a modest decrease in resistivity from just over 55 mΩ cm at 3 K to about 32 mΩ cm at 400 K, while a second crystal (crystal 2) showed a small increase in resistivity from about 8 mΩ cm at 80 K to just over 12 mΩ cm. The absence of a cusp in the temperature dependent resistivities in n-type samples may indicate a difference in coupling strength between the $Eu^{2+}$ magnetic spins and the majority carrier type. The cryostat data and PPMS data show good overlap and appear consistent over the measured temperature range. Overall, this resistivity range is in good agreement with previous reports and fits a moderately doped-semiconductor picture for $EuCd_2As_2$.[10, 14, 18, 21-24]

The carrier concentrations were extracted from the Hall resistivity curves, (Fig. S5), using a single carrier model. The temperature dependent carrier densities observed for all samples of $EuCd_2As_2$ are plotted in Figure 2c. The p-type carrier densities falling in a narrow range from $6.0 \times 10^{17}$ $cm^{-3}$ at low temperature to $8.3 \times 10^{17}$ $cm^{-3}$ at high temperature. The n-type carrier densities fall into similarly narrow ranges. Crystal 1 has a doping level of $1.7 \times 10^{17}$ $cm^{-3}$ at low temperatures that rises to $2.4 \times 10^{17}$ $cm^{-3}$ at high temperatures. Crystal 2 stays nearly constant across the measured temperature range at $6 \times 10^{17}$ $cm^{-3}$. These carrier densities are 1-3 orders of magnitude lower than the previous reports claiming semimetallic behavior[24, 30] and are consistent with $EuCd_2As_2$ as a doped semiconductor.

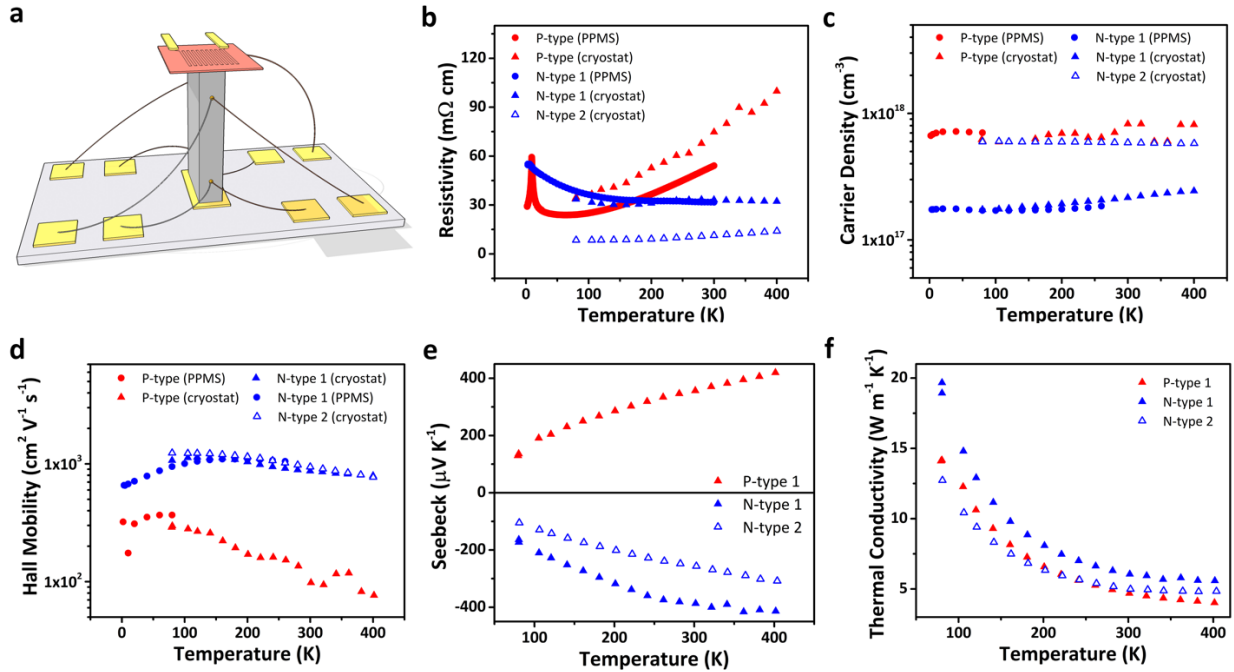

Figure 2. (a) Schematic diagram of a typical cryostat device used to measure the thermal and electrical transport properties of EuCd$_2$As$_2$ single crystals. The yellow slabs represent brass contacts and current spreaders. The red colored slab at the top represents the resistive heater used for establishing a thermal gradient. The gray vertical block in the middle represents the single crystal of EuCd$_2$As$_2$ and the light-gray baseplate represents the alumina base-plate. (b) – (f) Transport data measured on both p- and n-type EuCd$_2$As$_2$ single crystals. Hall data is extracted assuming single carrier transport. Circles represent PPMS data while triangles correspond to cryostat data. (b) Temperature dependent resistivity measured from <10 K to 400 K. (c) Temperature dependent carrier density measured from <10 K to 400 K. (d) Temperature dependent Hall mobility measured from <10 K to 400 K. (e) Temperature dependent thermopower measured from about 80 K to 400 K. (f) Temperature dependent thermal conductivity measured from about 80 K to 400 K.

Hall mobility was derived from the Hall coefficient and electrical resistivity. The temperature dependent Hall mobilities are plotted in Figure 2d. The p-type mobility is 322 cm$^2$ V$^{-1}$ s$^{-1}$ at 2 K and rises to 368 cm$^2$ V$^{-1}$ s$^{-1}$ at 80 K before finally falling to 76.5 cm$^2$ V$^{-1}$ s$^{-1}$ at about 400 K. The n-type mobility for crystal 1 is higher, beginning at 658 cm$^2$ V$^{-1}$ s$^{-1}$ at 3.5 K before reaching a maximum value of 1,144 cm$^2$ V$^{-1}$ s$^{-1}$ at 120 K. Finally, the n-type mobility drops to 796 cm$^2$ V$^{-1}$ s$^{-1}$ at 400 K. N-type crystal 2 has comparable mobilities. The fact that the n-type crystals have similar doping levels but higher mobilities than the p-type crystals, is consistent with the behavior in virtually all known semiconductors. Additionally, mobilities of this magnitude are typically more consistent with a semiconductor than a metal, excluding non-trivial topological effects that can substantially increase the mobility. The electron effective mass determined by the method of the four coefficients[48] and was found to be 0.07 m$_e$ at low temperature.

The temperature dependent thermopowers shown in Figure 2e were measured down to 80 K. The signs of the thermopower, along with the slopes of the Hall resistivity curves, allows unambiguous assignment of p- and n-type doping for the EuCd$_2$As$_2$ samples. The data are nearly symmetric about the zero line in the temperature dependent thermopower, with p-type thermopowers rising from 130 µV K$^{-1}$ at about 80 K to 420 µV K$^{-1}$ at about 400 K. The p-type thermopowers compare well with previous reports.[22][11] The n-type thermopower for crystal 1 on the other hand, drops from -164 µV K$^{-1}$ at around 80 K to about -415 µV K$^{-1}$ near 400 K. The thermopower values in the more highly doped crystal 2 have lower magnitudes, ranging from -105 to -308 µV K$^{-1}$ from 80 to 400 K, respectively.

Temperature dependent thermal conductivity was measured from 80 K to 400 K. The temperature dependent thermal conductivity of p-type and n-type EuCd$_2$As$_2$ have no significant differences. Both samples clearly show the expected features for a lattice-dominated thermal conductivity, namely the 1/T Umklapp phonon-phonon scattering trend with temperature. The p-type thermal conductivity at 80 K is about 14 W m$^{-1}$ K$^{-1}$ and drops off to about 4 W m$^{-1}$ K$^{-1}$ at 400 K. The n-type 80 K values are about 20 W m$^{-1}$ K$^{-1}$ falling to about 5.5 W m$^{-1}$ K$^{-1}$ at 400 K. These values are all within the errors in the measurements and agree well with previously reported thermal conductivities.[14]

The magnetic behavior of both p-type and n-type crystals was investigated by SQUID magnetometry to reveal any doping dependent differences in magnetism. Previous reports have demonstrated that the magnetic ground state of p-type EuCd$_2$As$_2$ is antiferromagnetic (AFM) with an in-plane A-type ordering (Figure 3a) and that in modest fields < 2 T EuCd$_2$As$_2$ can be fully ferromagnetically (FM) polarized.[10, 11, 21] Magnetization vs. applied field measurements at 2 K on both p-type and n-type crystals corroborate the AFM ground state, showing an absence of hysteresis, and both give polarization to the ferromagnetic state at a field of about 0.6 – 0.8 T for in-plane and about 1.3 – 1.7 T for cross-plane. The saturation moments from the magnetization data on p-type and n-type samples are 7.1 $\mu_B$ Eu$^{-1}$ and 7.6 $\mu_B$ Eu$^{-1}$, respectively. A previous study[13] evaluating EuCd$_2$As$_2$ crystals with different p-doping levels found that saturation moment is suppressed with increased p-doping. The increased saturation moment in our n-doped sample corroborates the observation that the saturation moment is quite sensitive to doping level. Magnetic susceptibility measurements from 2 K to 300 K further confirmed the AFM ground state of both p-type and n-type samples, which both have a TN of 9 K (Fig. 3c and 3d). The p-type ZFC susceptibility peaks slightly above 27 emu mol$^{-1}$ Oe$^{-1}$ and that of the FC susceptibility peaks just over 4 emu mol$^{-1}$ Oe$^{-1}$. The n-type magnetic susceptibility, however, remains practically identical from ZFC to FC, with both peaking right above 4 emu mol$^{-1}$ Oe$^{-1}$. A Curie-Weiss fit was applied to the temperature range from 11 K – 300 K for the inverse magnetic susceptibility yielding positive Curie-Weiss temperatures of 15.0 K and 9.9 K for p-type and n-type, respectively. The positive Curie-Weiss temperature is likely due to the strength of the ferromagnetic short-range interactions and has been seen previously in p-type $EuCd_2As_2$ as well as other similar $4f$ systems.[10, 11, 21, 49] The high temperature local saturation moment is also evident in the magnetic susceptibility, giving values of 7.9 $\mu_B$ Eu$^{-1}$ for the p-type $EuCd_2As_2$ and 7.4 $\mu_B$ Eu$^{-1}$ for the n-type. Further magnetic studies are required to fully understand the effect of La n-type doping on the magnetic properties of $EuCd_2As_2$.

To answer the question on whether $EuCd_2As_2$ is a topological semimetal or a trivial insulator, we performed angle-resolved photoemission spectroscopy measurements (ARPES) on n-type $EuCd_2As_2$ crystals. The $EuCd_2As_2$ bulk crystals are cleaved *in situ* and measured at T = 6 K. Figure 4a shows the energy dependent Fermi surface mapping along the A-Γ-A direction from 60 eV to 140 eV before potassium dosing. This allows for determining the photon energies that corresponded to $k_z = 0$ and $k_z = \pi/c$. Subsequently, 130 eV was selected when collecting further ARPES spectra as this approximately corresponds to $k_z = 0$. Figure 4b is the ARPES spectrum measured along the M-Γ-M direction on the 001-facet, while the K-Γ-K ARPES spectrum is shown in Fig. S6 for comparison.

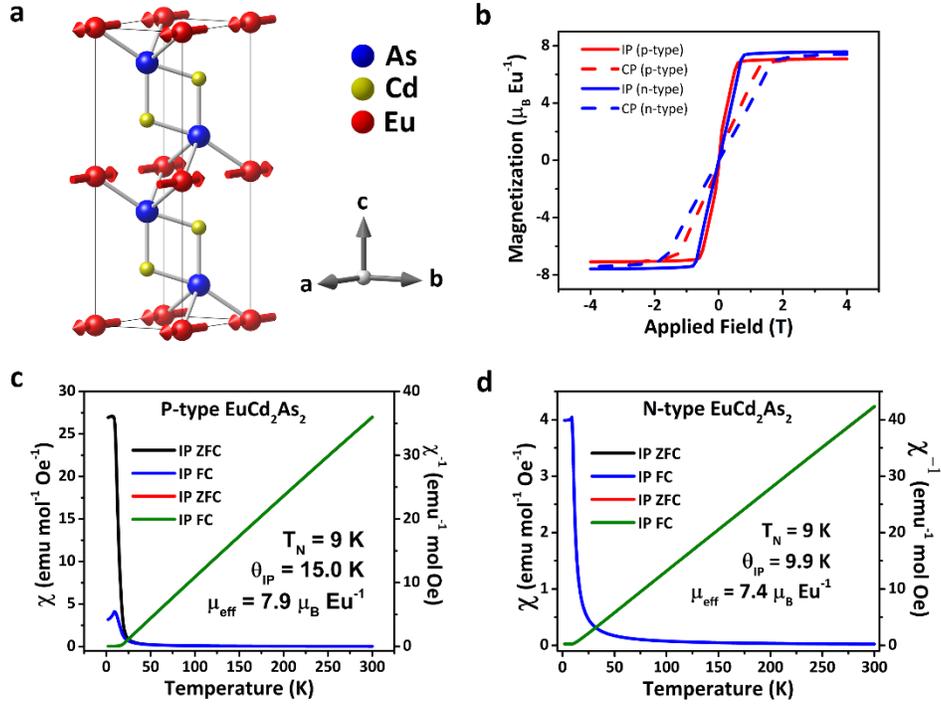

Figure 3. (a) Magnetic structure of antiferromagnetic $EuCd_2As_2$ showing in-plane ferromagnetic exchange and cross-plane anti-ferromagnetic exchange. (b) Field dependent magnetization of both p- and n-type $EuCd_2As_2$. (c) Temperature dependent magnetic susceptibility and inverse magnetic susceptibility of p-type $EuCd_2As_2$ measured from 2 K to 300 K. (d) Temperature dependent magnetic susceptibility and inverse magnetic susceptibility of n-type $EuCd_2As_2$ measured from 2 K to 300 K. The Curie-Weiss fitting was taken over the range from 11 K to 300 K for both p-type and n-type samples.

The VBM is located at the Γ point, with a cone-like E vs. k dispersion extending to about -0.8 eV, below the VBM. Between -0.8 and -2.0 eV, the photoemission spectrum is dominated by flat bands spreading across the entire Brillouin zone. These bands originate from Eu 4*f* electrons. All these features are consistent with previous ARPES studies on $EuCd_2As_2$ single crystals.[13, 18, 20] To directly visualize the conduction band, *in situ* potassium dosing was performed, as the combined transport signatures indicated that the Fermi levels of the moderately doped n-type crystals fell within the band gap. The ARPES spectrum after *in situ* potassium dosing is shown in

Figure 4c. It is clear that additional photoemission signal appears at the Γ point, with maximum intensity located at about +0.8 eV above the VBM. The energy dependent Fermi surface mapping was not repeated after potassium dosing, however all bands below 0 eV in Figures 4b and 4c are essentially superimposable to one another, which implies negligible shifts in $k_x$, $k_y$, or $k_z$ following dosing. Given that there are no other bands observed between the VBM and this new signal at +0.8 eV the new band must be the conduction band minimum (CBM), which is only visible when there is sufficient population of electrons in the conduction band. While a topological semimetallic electronic structure would also result in the appearance of a gap between a lower energy CBM and a higher energy VBM (Fig. 1a), an 800 meV semimetallic band overlap is over an order of magnitude larger than the overlap predicted for $EuCd_2As_2$ (~50 meV). [12, 18, 19] Furthermore, calculations predicted the nodal crossings to occur at a kz with 0.1 Å$^{-1}$ of the Γ point.[19] Together this indicates that the n-type $EuCd_2As_2$ is a magnetic insulator instead of a topological semimetal. Figure 4d plots the energy distribution curves before (red) and after (blue) potassium dosing, giving the density of states as a function of energy. Fitting these curves at the Γ point using Lorentzian line shapes yields a band gap of ~0.8 eV for our $EuCd_2As_2$ samples with an effective electron concentration $n_e$ = 6 x 10$^{18}$ cm$^{-3}$ after potassium dosing (Fig. S7). To corroborate the band gap, DRIFTS measurements were collected on powders of both p-type and n-type $EuCd_2As_2$, and the Tauc-Davis-Mott relations were applied assuming a three-dimensional density of states. Figures 4e and 4f are the Tauc plots for p-type and n-type $EuCd_2As_2$, respectively. The band edges are clearly visible and extrapolated to 0.72 eV for p-type and 0.71 eV for n-type, in good agreement with the ARPES gap considering the DRIFTS data was collected at room temperature while the ARPES spectrum was collected at 6 K. The reflectance data is also in excellent agreement with a previous report showing an absorbance edge at about 0.74 eV.[29]

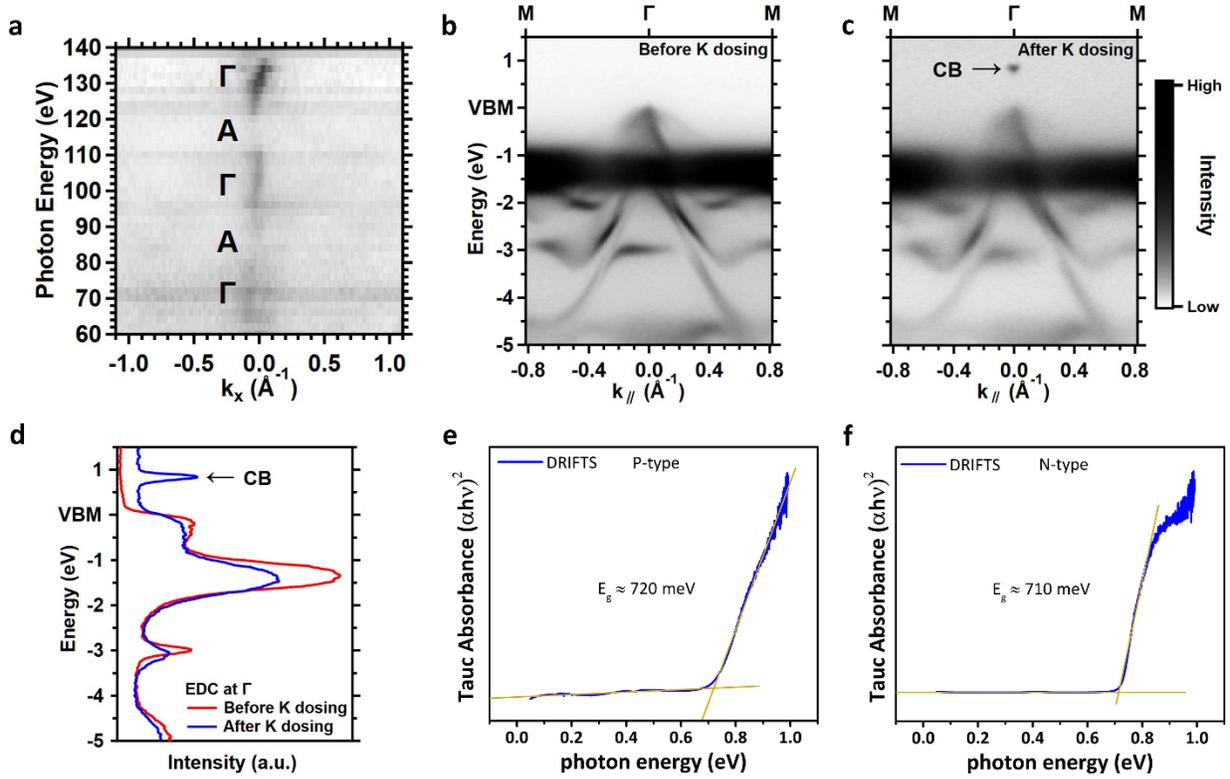

Figure 4. (a) Energy dependent Fermi surface mapping along the cross-plane A-Γ-A direction ($k_z$-axis) from 60 eV to 140 eV. The $k_x$-axis lies in the M-Γ-M in-plane direction. 130 eV was selected to access the $k_z = 0$ plane. (b) Intensity plot for angle resolved photoemission spectroscopy (ARPES) measured at 6 K on an n-type sample of $EuCd_2As_2$ showing the dispersion over the M-Γ-M direction on a cleaved 001-facet. (c) ARPES intensity plot measured on the same n-type sample at 6 K after dosing the surface with evaporated potassium metal (K-dosing). Note the bottom of the conduction band is clearly visible at ~800 meV above the top of the valence band. (d) Energy distribution curves showing the energy dependent density of states at Γ. Note the clear emergence of conduction band states at about 800 meV after K-dosing. (e) Diffuse reflectance infrared Fourier transform spectroscopy (DRIFTS) Tauc plot measured on p-type $EuCd_2As_2$ assuming an allowed 3D direct transition. (f) DRIFTS Tauc plot measured on n-type $EuCd_2As_2$ assuming an allowed 3D direct transition.

CONCLUSION

These results directly establish the semiconducting nature of EuCd$_2$As$_2$ with a 0.7-0.8 eV band gap, through the growth and characterization of n-type EuCd$_2$As$_2$ crystals. It was demonstrated that incorporation of La into the Sn-flux synthesis as a dopant led to n-type thermopower and Hall effect signatures. The conduction band minimum was observed 0.8 eV above the valence band maximum at 6 K through ARPES measurements, with no other states observed within the gap. Room temperature diffuse reflectance absorbance measurements further confirmed a 0.71 – 0.72 eV optical band gap. Together, this corroborates the recent reports of semiconducting behavior, and with such a large band gap, precludes the possibility of realizing Weyl semimetallic behavior in EuCd$_2$As$_2$ under most accessible experimental conditions. These results indicate that the realization of magnetic Weyl semimetals, and the scientifically and technologically intriguing phenomena that they may exhibit, will require further materials exploration, and much more grounded expectations.

ASSOCIATED CONTENT

**Supporting Information**. Included in the Supporting Information is the predicted Weyl semimetal EuCd$_2$As$_2$ band structure, the ToF-SIMS data, Rietveld refinements of both p- and n-type EuCd$_2$As$_2$, Hall resistivity curves, and additional ARPES data.

AUTHOR INFORMATION

**Corresponding Author**

*E-mail: goldberger@chemistry.ohio-state.edu

**Author Contributions**

The manuscript was written through contributions of all authors. All authors have given approval to the final version of the manuscript.

ACKNOWLEDGMENT

This work was primarily supported by the ONR MURI "Extraordinary electronic switching of thermal transport", grant number N00014-21-1-2377, for JEG, RAN, NRJ, JCK, LS, JPH, EM, and CC. This research was partially supported by the Center for Emergent Materials, an NSF MRSEC, under award number DMR-2011876 for AJW, JK, RK, and SC. JK also acknowledges funding from the U.S. Department Office of Science, Office of Basic Sciences, of the U.S. Department of Energy under Award No. DE-SC0020323. This research used resources of the Advanced Light Source, which is a DOE Office of Science User Facility under contract no. DE-AC02-05CH11231.

**Supplementary Information for:**

# Revealing the EuCd$_2$As$_2$ Semiconducting Band Gap via n-type La-Doping

*Ryan A. Nelson[1], Jesaiah King[1], Shuyu Cheng[2], Archibald J. Williams[1], Christopher Jozwiak[8], Aaron Bostwick[8], Eli Rotenberg[8], Souvik Sasmal[5], I-Hsuan Kao[5], Aalok Tiwari[5], Natalie R. Jones[1], Chuting Cai[6], Emma Martin[7], Andrei Dolocan[9], Li Shi[6,7], Roland Kawakami[2], Joseph P. Heremans[2-4], Jyoti Katoch[5], Joshua E. Goldberger[1,4]\**

[1]Department of Chemistry and Biochemistry, The Ohio State University, Columbus, Ohio 43210, USA

[2]Department of Physics, The Ohio State University, Columbus, Ohio 43210, USA

[3]Department of Mechanical and Aerospace Engineering, The Ohio State University, Columbus, Ohio 43210, USA

[4]Department of Materials Science and Engineering, The Ohio State University, Columbus, Ohio 43210, USA

[5]Department of Physics, Carnegie Mellon University, Pittsburgh, PA, 15213, USA

[6]Materials Science and Engineering Program, The University of Texas at Austin, Austin, Texas 78712, USA

[7]Department of Mechanical Engineering, The University of Texas at Austin, Austin, Texas 78712, USA

[8]Advanced Light Source, E.O. Lawrence Berkeley National Laboratory, Berkeley, California 94720, USA

⁹Texas Materials Institute, The University of Texas at Austin, Austin, Texas 78712, USA

**EuCd$_2$As$_2$ Band Structure**

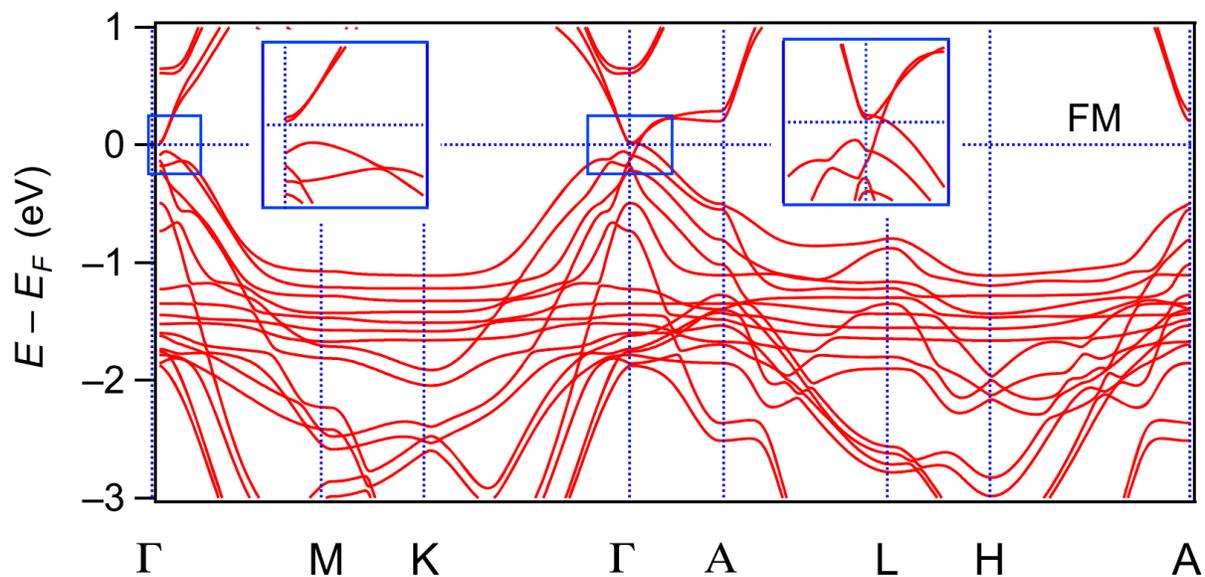

**Fig. S1** EuCd$_2$As$_2$ band structure in the ferromagnetic phase along high-symmetry lines calculated by using DFT + U, with U = 5 eV. The magnetic moments oriented in the (001) direction. The insets are the zoomed-in views of band dispersions in the vicinity of the E$_F$ around the Γ point. Band structure cropped from original publication by Tian Qian, Ming Shi, and Y.G. Shi, used under CC BY-NC 4.0.[1]

## Time-of-flight secondary ion mass spectroscopy (ToF-SIMS)

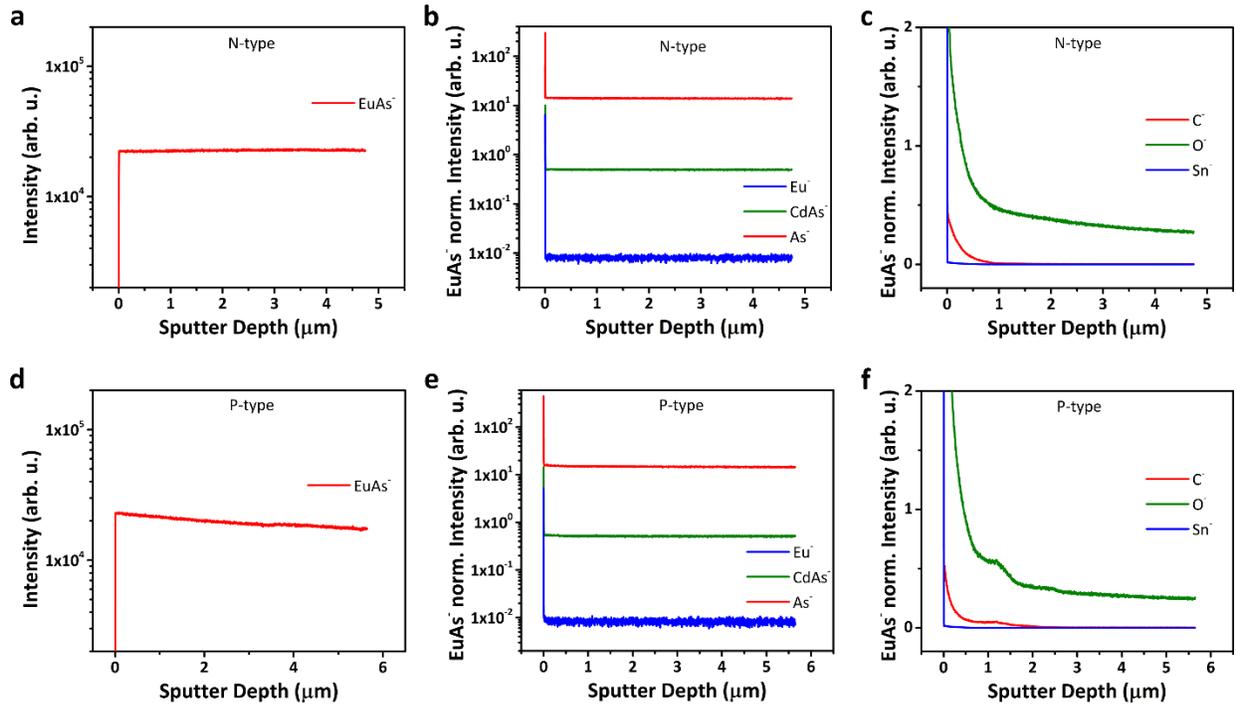

**Fig. S2** Time-of-flight secondary ion mass spectroscopy in negative polarity mode. (a) Raw ion intensity profile for the EuAs$^-$ ion as a function of sputter depth in the n-type EuCd$_2$As$_2$. (b) Normalized ion intensity profiles for Eu$^-$, CdAs$^-$, and As$^-$ ions as a function of sputter depth in the n-type EuCd$_2$As$_2$. Raw ion intensities were normalized to the EuAs$^-$ ion intensity profile in (a). (c) Normalized ion intensity profiles for C$^-$, O$^-$, and Sn$^-$ ions as a function of sputter depth in the n-type EuCd$_2$As$_2$. Raw ion intensities were normalized to the EuAs$^-$ ion intensity profile in (a). (d) Raw ion intensity profile for the EuAs$^-$ ion as a function of sputter depth in the p-type EuCd$_2$As$_2$. (e) Normalized ion intensity profiles for Eu$^-$, CdAs$^-$, and As$^-$ ions as a function of sputter depth in the p-type EuCd$_2$As$_2$. Raw ion intensities were normalized to the EuAs$^-$ ion intensity profile in (d). (f) Normalized ion intensity profiles for C$^-$, O$^-$, and Sn$^-$ ions as a function of sputter depth in the p-type EuCd$_2$As$_2$. Raw ion intensities were normalized to the EuAs$^-$ ion intensity profile in (d).

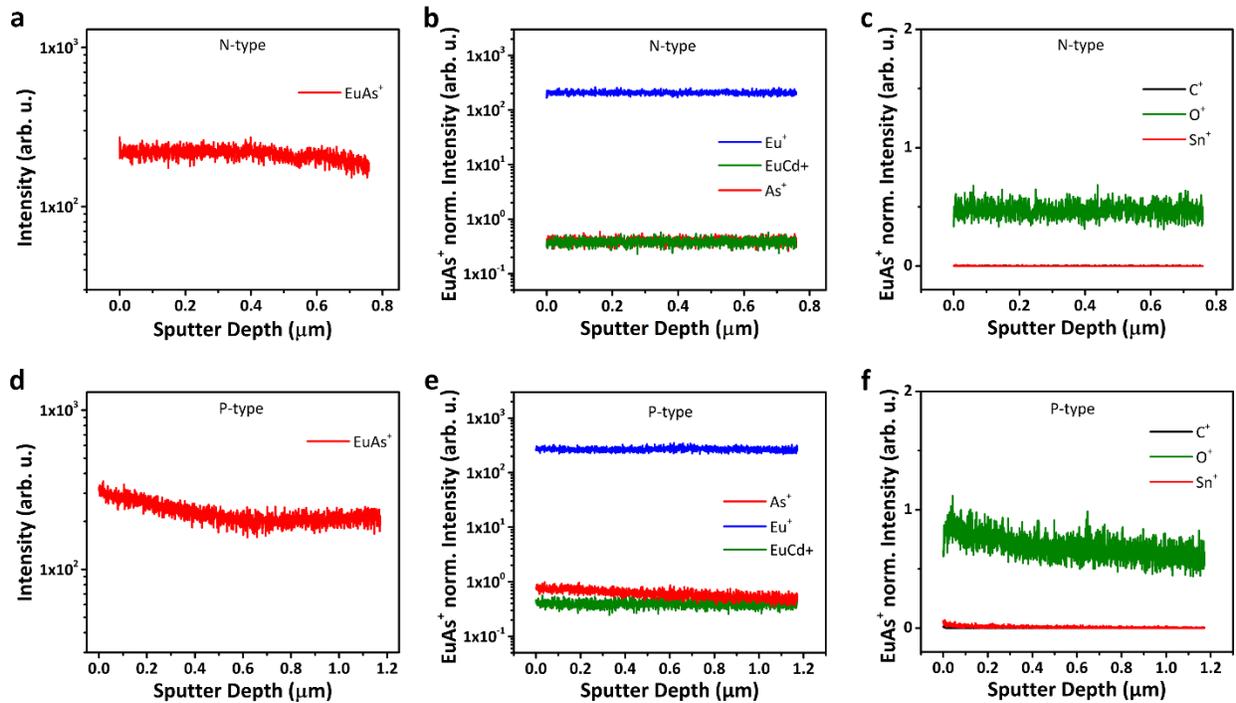

**Fig. S3** Time-of-flight secondary ion mass spectroscopy in positive polarity mode. (a) Raw ion intensity profile for the EuAs⁻ ion as a function of sputter depth in the n-type EuCd$_2$As$_2$. (b) Normalized ion intensity profiles for Eu⁻, EuCd⁻, and As⁻ ions as a function of sputter depth in the n-type EuCd$_2$As$_2$. Raw ion intensities were normalized to the EuAs⁻ ion intensity profile in (a). (c) Normalized ion intensity profiles for C⁻, O⁻, and Sn⁻ ions as a function of sputter depth in the n-type EuCd$_2$As$_2$. Raw ion intensities were normalized to the EuAs⁻ ion intensity profile in (a). (d) Raw ion intensity profile for the EuAs⁻ ion as a function of sputter depth in the p-type EuCd$_2$As$_2$. (e) Normalized ion intensity profiles for Eu⁻, EuCd⁻, and As⁻ ions as a function of sputter depth in the p-type EuCd$_2$As$_2$. Raw ion intensities were normalized to the EuAs⁻ ion intensity profile in (d). (f) Normalized ion intensity profiles for C⁻, O⁻, and Sn⁻ ions as a function of sputter depth in the p-type EuCd$_2$As$_2$. Raw ion intensities were normalized to the EuAs⁻ ion intensity profile in (d).

The ToF-SIMS data in Figures S2 and S3 show that there is no significant compositional difference discernable between the n-type and p-type samples. The ions that come from the lattice elements (i.e. Eu, Cd, As containing ions) do not vary with depth, showing that they are the lattice components. Further, this is an indication that the majority n-type and p-type dopants are below the detection limits.

The impurity elements besides oxygen are largely located on the surface of the crystals as is indicated by the rapid drop to zero intensity (most apparent in Fig. S2c and f). It is clear that there is some small level of oxygen present in both p-type and n-type samples, nevertheless, the oxygen impurity levels are very similar between the n-type and p-type samples and does not appear to have a significant impact on the majority dopants. The C, O, and Sn ions are a representative sample of other surface and low-level contaminates that were identified, but similarly found to show negligible differences between the two samples. Nearly all contaminates identified were surface contaminates and were sputtered away early into the measurements.

## X-ray diffraction Rietveld refinements

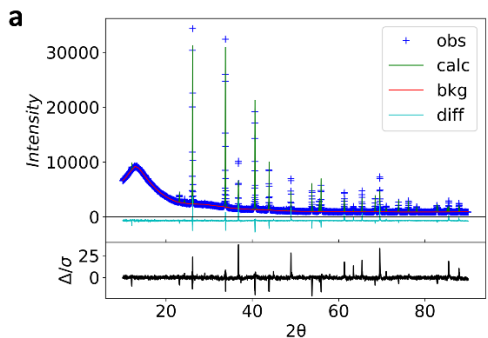

**p-type EuCd$_2$As$_2$**

| | |
|---|---|
| a (Å) | 4.43965(9) |
| c (Å) | 7.32665(8) |
| Eu position | 0, 0, 0 |
| Cd position | 1/3, 2/3, 0.6347(3) |
| As position | 1/3, 2/3, 0.2516(5) |
| wR | 5.21% |

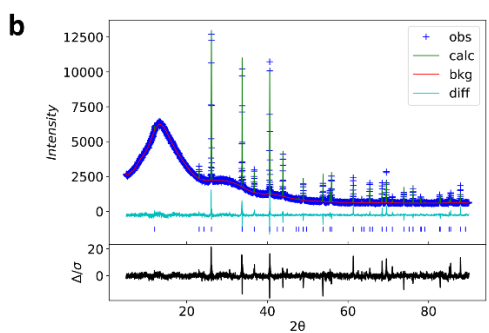

**n-type EuCd$_2$As$_2$**

| | |
|---|---|
| a (Å) | 4.44016(11) |
| c (Å) | 7.32667(11) |
| Eu position | 0, 0, 0 |
| Cd position | 1/3, 2/3, 0.6294(4) |
| As position | 1/3, 2/3, 0.2541(6) |
| wR | 4.18% |

**Fig. S4** GSAS II Rietveld refinements for (a) p-type EuCd$_2$As$_2$ and (b) n-type EuCd$_2$As$_2$ x-ray diffraction powder patterns.

We employed GSAS II for Rietveld refinements[2] of the lattice parameters and atom positions to compare them with each other and with the previously reported values. N-type doping seems to have a negligible influence on the crystal lattice as it refines to be nearly identical to the p-type data and to the ICSD reference pattern.[3] The x-ray data further shows that there is no significant structural difference between the p-type and n-type material.

## Hall effect resistivity curves

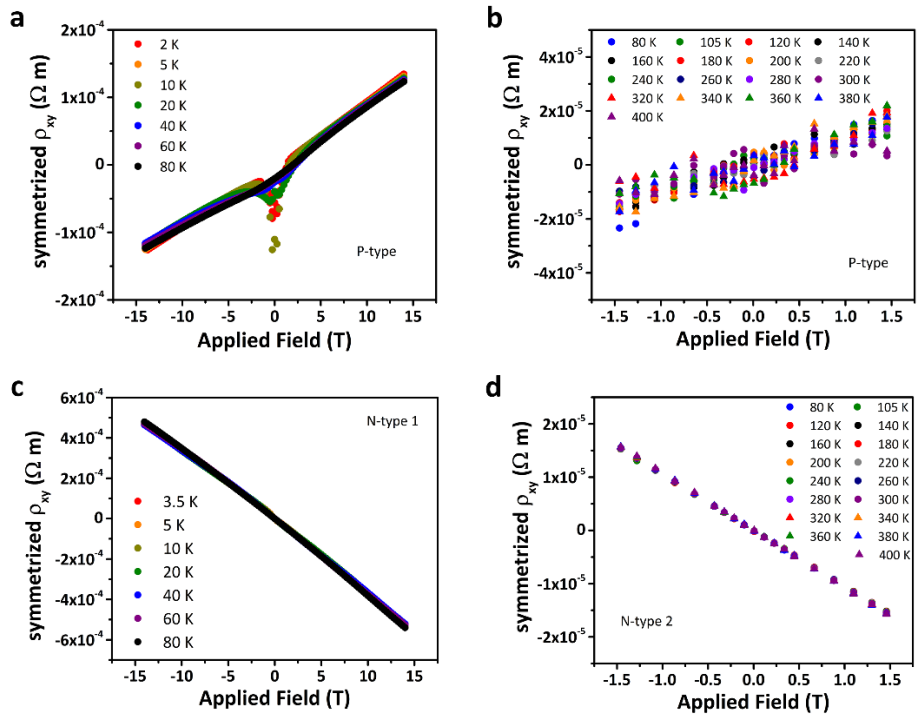

**Fig. S5** (a) Low temperature symmetrized Hall effect resistivity curves measured on the p-type sample from the main text using a 14 T PPMS. Note the anomalous Hall signal at low fields. (b) High temperature symmetrized Hall effect resistivity curves measured on the same p-type sample using a 1.4 T cryostat. (c) Low temperature symmetrized Hall effect resistivity curves measured on n-type sample 1 from the main text using the PPMS. Note the absence of any significant anomalous Hall signal. (d) High temperature symmetrized Hall effect resistivity curves measured on n-type sample 2 using the cryostat.

## Angle-resolved photoemission spectroscopy

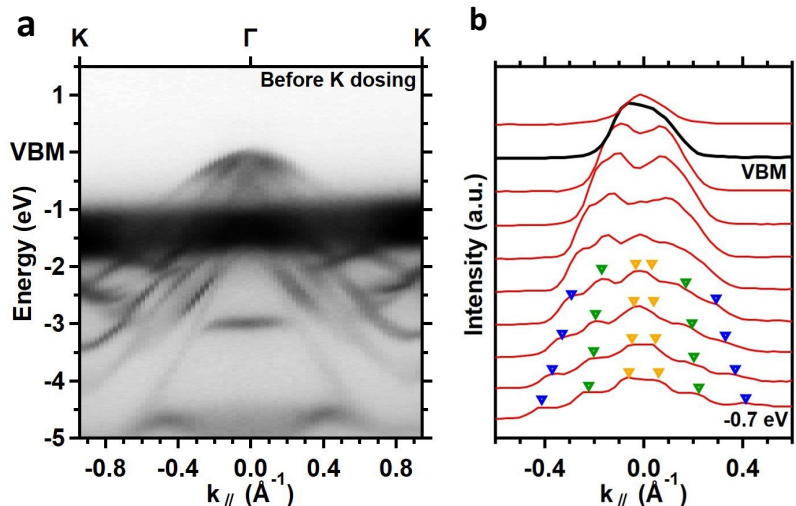

**Fig. S6** (a) ARPES spectrum along the K-Γ-K direction before K dosing. (b) Representative momentum distribution curves (MDCs) near the VBM at the Γ point. The positions of the MDC peaks are marked by the triangles.

Fig. S6a shows the ARPES spectrum along the K-Γ-K direction before potassium dosing. Near the VBM, three sets of cone-like bands can be identified from the spectrum along the K-Γ-K direction. This is visualized more clearly by the representative momentum distribution curves (MDCs) near the VBM, as shown in Fig. S6b. The peak positions of the MDCs are extracted using Lorentzian fitting, under the assumption that the peak positions are symmetric about the Γ point. The fitted MDC peak positions marked by the orange, green, and blue triangles in Fig S6b. The dark bands between -1 eV and -2 eV that span the entire spectrum in Fig. s6a are the 4f states.

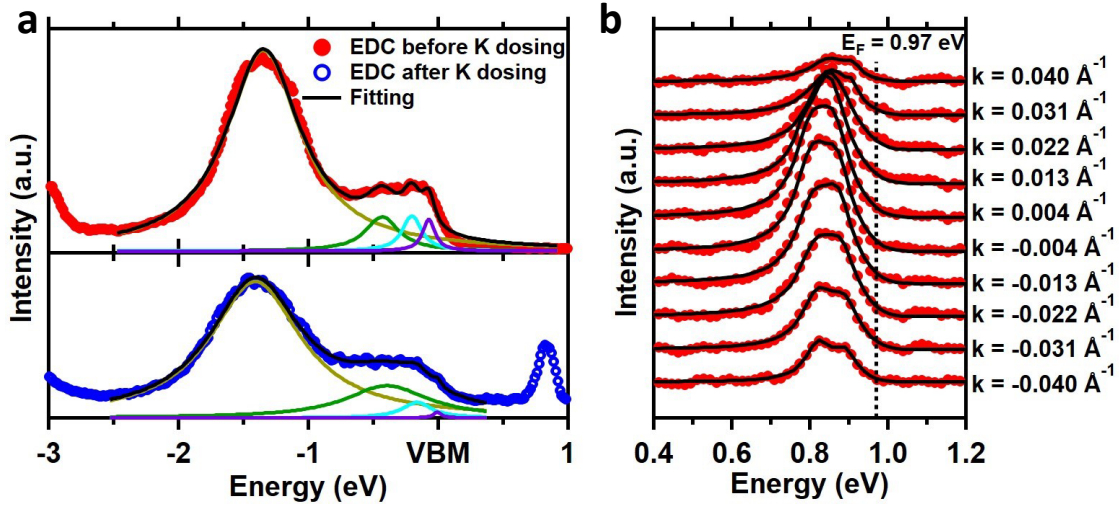

**Fig. S7** Lorentzian fitting of EDCs. a) EDCs the Γ point before (red) and after (blue) K dosing. The black curves are the fitting results, and the yellow, green, cyan and purple curves are individual Lorentzian components. b) Several representative EDCs around the Γ point at the CBM. The red dots are the experimental data, the black curves are the fitting results, and the vertical dashed line shows the position of the Fermi level.

We fit the energy distribution curves (EDCs) of the VB using four Lorentzian components, three from the cone-like bands near the VBM (the green, teal and blue curves in Fig. S7a, and one from the Eu 4f band at -1.4 eV (the yellow curve in Fig. S7a). The experimentally measured EDCs at the Γ point before and after potassium dosing are displayed in red and blue dots in Fig. S7a, respectively. The black curves in Fig. S7a show the fitting results.

The position of the CB is determined by fitting several representative EDCs around the Γ point, as shown in Fig. S7b. Double-peak features can be seen in the EDCs at the CBM. Therefore, the function used to fit the EDCs at the CBM is the product of Fermi-Dirac distribution function and the sum of two Lorentzian components, convolved with a Gaussian function representing instrumental energy resolution. The Fermi level position is determined by fitting the EDCs at $k = \pm 0.040$ Å$^{-1}$, which gives consistent results of $E_F = 0.97\ eV$ above the VBM. Fitting the EDCs yields a CBM position of 0.80 eV above the VBM. Therefore, the band gap of our EuCd$_2$As$_2$ is 800 meV.

The surface potassium dosage is estimated following the methods described by Katoch et al.[4] Since the band structure of EuCd$_2$As$_2$ is three-dimensional, the effective carrier concentration can be estimated by:

$$n_e = \frac{g}{4\pi^2}\left(\frac{2m^*k_BT}{\hbar^2}\right)^{3/2}\int_0^{\frac{E_F-E}{k_BT}}\frac{\sqrt{u}}{1+\exp{(u-\frac{E_F-E}{k_BT})}}du$$

where $g$ is the degeneracy of the bands, $m^*$ is the electron effective mass at the CBM, $E_F$ is the Fermi level. Here we use $g = 2$ in our estimation since we observe double-peak features in EDCs at the VBM. The effective mass can be derived from the method of the four coefficients as shown by Jovovic et al.[5] The effective mass at low temperature was found to be $m^* = 0.07\, m_0$ based on thermoelectric transport measurements. As previously mentioned above, $E_F - E$ is found to be 0.17 eV. Substituting in all these values into the above equation yields an effective electron concentration of $n_e = 6 \times 10^{18}\, cm^{-3}$.